\documentclass[twocolumn,useAMS,usenatbib,usegraphicx]{mn2e}
\usepackage[utf8]{inputenc}
\usepackage[UKenglish]{babel}
\usepackage{url}
\usepackage{amssymb}
\usepackage{aas_macros}

\newcommand{\rnl}{\mathcal{R}(f_{\rm NL})}

\newcommand{\lcdm}{$\Lambda$CDM }

\newcommand{\fsky}{f_{\rm sky}}

\newcommand{\cardiff}{{School of Physics \& Astronomy, Cardiff University, 5 The Parade, Cardiff, CF24 3AA, United Kingdom}}

\pagerange{L19--L23}
\volume{421}
\pubyear{2012}
\date{Accepted 2011 November 25. Received 2011 November 25; in original form 2011 November 4}

\title{Testing Cosmology with Extreme Galaxy Clusters}
\author[Harrison \& Coles]{Ian Harrison\thanks{E-mail: ian.harrison@astro.cf.ac.uk} and Peter Coles \\ \cardiff}

\begin{document}
\maketitle
\begin{abstract}
Motivated by recent suggestions that a number of observed galaxy
clusters have masses which are too high for their given redshift to
occur naturally in a standard model cosmology, we use Extreme Value
Statistics to construct confidence regions in the mass-redshift
plane for the most extreme objects expected in the universe. We show
how such a diagram not only provides a way of potentially ruling out
the concordance cosmology, but also allows us to differentiate
between alternative models of enhanced structure formation. We
compare our theoretical prediction with observations, placing
currently observed high and low redshift clusters on a mass-redshift
diagram and find -- provided we consider the full sky to avoid
\textit{a posteriori} selection effects -- that none are in
significant tension with concordance cosmology.
\end{abstract}

\section{Introduction}
In the standard ``concordance'' ($\Lambda$CDM) model of cosmology, structure
formation proceeds in a hierarchical, `bottom-up' fashion, with
small scale perturbations in the initial distribution of Cold Dark
Matter (CDM) collapsing first, before merging over time to form
larger and larger haloes. The initial, nearly Gaussian,
distribution of these overdensities means that large fluctuations,
and correspondingly large halo masses, should be extremely rare in
the early universe. However, many of the plausible extensions to the
concordance model have been found to be capable of enhancing (or
depleting) structure formation, including primordial non-Gaussianity
\citep{Lucchin1988a, Pillepich2010a}, modified gravity
\citep{Schmidt2009, Ferraro2011} and scalar field \citep{Baldi2011c, Mortonson2011, Tarrant2011} scenarios. Furthermore, each of
these extensions can affect structure formation in different ways at
different times in the history of the universe, meaning that if we
can construct a history of the growth of structure we can
discriminate between competing models.

In addition, much attention has recently been paid to the prospect
that the existence of a single extreme (in terms of its high mass
\emph{and} redshift) object in the universe has the potential to
rule out at a high significance level cosmological models in which
it is correspondingly unlikely to exist. Following observations of a
series of apparently extreme objects \citep{Jee2009a, Brodwin2010,
Foley2011, Menanteau2011, PlanckCollaboration2011, Santos2011}, some
authors have claimed that such objects are either highly unlikely to
exist \citep{Jimenez2009, Holz2010, Cay'on2011, Hoyle2011b} in a
concordance \lcdm cosmology or, more powerfully, are significantly
larger than the expected \emph{most massive} object in a \lcdm
universe \citep {Chongchitnan2011}, although \cite{Waizmann2011a}
contend this result. As well as highlighting tensions with the
standard model, various authors have shown how such high-mass,
high-redshift objects may be explained by an enhanced rate of
structure formation generated by the inclusion of primordial
non-Gaussianity \citep{Enqvist2011, Hoyle2011b, Cay'on2011,
Chongchitnan2011} or scalar field dark energy \citep{Waizmann2011}
into the concordance cosmology.

In this paper we build on our previous work using extreme value
statistics \citep{Harrison2011} to test the current
concordance model with extreme galaxy clusters as well investigating
the different behaviours of extreme clusters in alternative models.
We construct the probability contours of the highest mass cluster
for all redshifts in an observational survey and directly compare
the current best fit \lcdm model with observations showing that, if
we consider the full sky (in order to avoid \textit{a posteriori}
selection effects) then none of the currently observed extreme
galaxy clusters are in tension with the \lcdm model. Then, using
information from the CoDECS \citep{Baldi2011a} N-body simulations,
we show how, if any future observations do exclude the \lcdm model,
extreme clusters can potentially be used to understand which of the
alternative models available may best explain the enhanced structure
formation.

The paper is organised as follows. In section \ref{sec:methods} we
introduce extreme value statistics and show how they may be used to
predict the mass of the most extreme cluster within an observational
survey. Section \ref{sec:lcdm} compares predictions for a \lcdm
universe with observations. Section \ref{sec:codecs} shows how the plot of extreme value
contours against redshift can highlight differences between cosmological models. In section \ref{sec:conclusions} we conclude and discuss
prospects for future work in this area.
\section{Method}
\label{sec:methods}
Extreme Value Statistics (EVS) \citep{Gumbel, KatzEVD} seeks to make
predictions for the properties of the greatest (or least) valued
random variable drawn from an underlying distribution. If we
consider a sequence of $N$ random variates $\lbrace M_i \rbrace$
drawn from a cumulative distribution $F(m)$ then there will be a
largest value of the sequence: $M_{\rm max} \equiv \sup \lbrace M_1, \ldots
M_N \rbrace.$
If these variables are mutually independent and identically
distributed then the probability that all of the deviates are less
than or equal to some $m$ is given by:
\begin{eqnarray}
\label{eqn:evs:evs_cdf} \Phi(M_{\rm max} \leq m;N) & = & F_1(M_1
\leq m)\ldots F_N(M_N \leq m)\nonumber\\
 & = & F^{N}(m)
\end{eqnarray}
and the probability density function (pdf) for $M_{\rm max}$ is then
found by differentiating (\ref{eqn:evs:evs_cdf}):
\begin{equation}
\label{eqn:evs:evs_exact} \phi(M_{\rm max} = m;N) = N f(m) \left[
F(m) \right]^{N-1},
\end{equation}
where $f(m)=dF(m)/dm$ is the pdf of the original distribution. This
gives the \emph{exact} extreme value pdf for $N$ observations drawn
from a known distribution $F(m)$; for more discussion of the
advantages of this approach over others involving the use of
asymptotic theory, see \\cite{Harrison2011}. To apply this general result to the concrete example of the most
massive cluster we use the appropriate halo mass function, which gives
the number density of haloes $n(M)$, to derive $f(m)$ and $F(m)$
according to:
\begin{eqnarray}
\label{eqn:hmf_evs:evs_exact}
f(m) &=& \frac{1}{n_{\rm tot}}\frac{dn(m)}{dm}, \\
F(m) &=& \frac{1}{n_{\rm tot}}\left[ \int_{-\infty}^{m} dM \,
\frac{dn(M)}{dM} \right],
\end{eqnarray}
where the normalisation factor
\begin{eqnarray}
\label{eqn:evs_exact:norm} n_{\rm tot} = \int_{-\infty}^{\infty} dM
\, \frac{dn(M)}{dM}
\end{eqnarray}
is the total (co-moving) number density of haloes. For a constant
redshift box of volume $V$ the total number of expected haloes $N$
is given by $n_{\rm tot}V$. These distributions can be inserted into
equation (\ref{eqn:evs:evs_exact}) to predict the pdf of the highest
mass dark matter halo within the volume. Although this procedure
explicitly assumes that haloes are uncorrelated, we have found
\citep{Harrison2011} that the results closely match those of
\cite{Davis2011}, who construct their extreme value distribution as
the differential of the void probability:
\begin{eqnarray}
\label{eqn:hmf_evs:ddcsp_evt} \Phi^{\rm void}(M_{\rm max} = m) =
\frac{dP_{0}(m)}{dm},
\end{eqnarray}
(allowing them to account for complications due to halo correlations
and bias), for the high masses of relevance to the inference we are
interested in.

In a cosmological survey we observe clusters at various redshifts
along our past light cone rather than on a single spatial
hypersurface at fixed $z$. If we wish to construct the EVS for
galaxy clusters within an observational survey which covers a
fraction $\fsky$ of the sky between redshifts $z_{\rm min}$ and
$z_{\rm max}$ we therefore need to take into account both the effect
of the growth of structure with decreasing redshift on the halo mass
function $n(m,z)$ and the observational volume we are probing in an
expanding universe, via the volume element $dV/dz$. Doing this
allows us to form the pdf of halo masses within that survey as:
\begin{eqnarray}
        \label{eqn:obs_pdf}
        f(m) &=& \frac{\fsky}{N_{tot}}\left[ \int^{z_{\rm max}}_{z_{min}} dz \, \frac{dV}{dz}\frac{dn(m, z)}{dm} \right], \\
        F(m) &=& \frac{\fsky}{N_{tot}}\left[ \int^{z_{\rm max}}_{z_{min}} \int^{m}_{-\infty} dz \, dM \, \frac{dV}{dz}\frac{dn(M, z)}{dM} \right],
\end{eqnarray}
where
\begin{eqnarray}
        \label{eqn:obs_norm}
        N_{tot} &=& \fsky \left[ \int^{z_{\rm max}}_{z_{min}} \int^{\infty}_{-\infty} dz \, dM \, \frac{dV}{dz}\frac{dn(M, z)}{dM} \right].
\end{eqnarray}
and then feed these distributions into our extreme value
prescription (\ref{eqn:evs:evs_exact}) (of course it is impractical to integrate numerically to infinite endpoints and so finite limits of $12 < \log_{10}m < 18$ are chosen; we have checked that this choice makes no difference to the conclusions). In order to make best use of
this information, we want to be able to see the distributions for
all redshifts at once; we hence construct the EVS distribution for
narrow bins in redshift space $\Delta z = 0.02$ (chosen so that $N_{bins} >> N_{clusters}$ and the highest expected mass for all redshifts remains the same as for $N_{bins}=1$), integrate over
these pdfs to find the $66\%, 95\%$ and $99\%$ confidence regions
and plot these, along with the peak of the distribution, for all
redshifts $0 < z < 2$. This can then be used to test the
cosmological model: if an observed cluster lies above e.g. the
$95\%$ region of such a distribution, then we may say we have a
correspondingly significant detection of enhanced structure
formation at that redshift.
\section{Comparison with Observations}
\label{sec:lcdm} We can now apply this technique to find out if any
currently observed clusters are discordant with the concordance
\lcdm predictions. We emphasize that, because we are predicting the
distributions of the \textit{most massive} cluster at each redshift,
if even a single galaxy cluster lying outside the extreme value
contours when placed on a mass-redshift plot can be seen as a
significant detection of deviation from concordance cosmology.
\subsection{Calibration of Cosmology and Cluster Masses}
\begin{table}
\label{tab:clusters}
\begin{center}
\begin{tabular}{lcc} \hline
Cluster & $z$ & $M_{200\rm m}^{\rm Edd}/M_\odot$  \\
\hline
A2163$^{1}$ & $0.203$ & $3.04_{-0.67}^{+0.87}\times 10^{15}$  \\
A370$^{1}$ & $0.375$ & $2.62_{-0.67}^{+0.87}\times 10^{15}$ \\
RXJ1347$^{1}$ & $0.451$ & $2.14_{-0.48}^{+0.60}\times 10^{15}$  \\
ACT-CL~J0102$^{2}$  & $0.87$ & $1.85_{-0.33}^{+0.42}\times 10^{15}$ \\
PLCK~G266$^{3}$ & $0.94$ & $1.45^{+0.27}_{-0.20}\times 10^{15}$  \\
SPT-CL~J2106$^{4}$  & $1.132$ & $1.11_{-0.20}^{+0.24}\times 10^{15}$  \\
SPT-CL~J0546$^{5}$ & $1.067$ & $7.80^{+1.27}_{-0.90}\times 10^{14}$ \\
XXMU~J2235$^{6}$  & $1.4$ & $6.82_{-1.23}^{+1.52}\times 10^{14}$   \\
XXMU~J0044$^{7}$  & $1.579$ & $4.02_{-0.73}^{+0.88}\times 10^{14}$  \\
\hline
\end{tabular}
\end{center}
\caption{The extreme clusters considered in this paper ([1]
\citealt{Maughan2011}, [2] \citealt{Menanteau2011},
[3]\citealt{PlanckCollaboration2011}, [4] \citealt{Foley2011}, [5]
\citealt{Brodwin2010}, [6] \citealt{Jee2009a}, [7]
\citealt{Santos2011}). $M_{200\rm m}^{\rm Edd}$ is calculated using
the numerical code of \citep{Zhao2009} to convert from $M_{200\rm
c}$ (where necessary) and equation (\ref{eqn:edd}) to include the
Eddington bias.}
\end{table}
In order to meaningfully compare our theoretical predictions to
observations, we need to carefully ensure our \lcdm cosmology is
correctly calibrated. As the ingredients for our concordance
cosmology here, we use a linear matter power spectrum $P(k)$
calculated using the numerical code
CAMB\footnote{\url{http://camb.info}} and the WMAP7+BAO+H0 Mean
parameters from \cite{Komatsu2011n}. From this we calculate the
variance of the matter field, smoothed with a top hat window $W(k;
R)$ of radius $R=(3m/4\pi\bar{\rho}_{m,0})^{1/3}$, evolved to a
redshift $z$ with the linear growth function $D_{+}(z)$ (normalised
to $D_{+}(0)=1$):
\begin{eqnarray}
\label{eqn:sigsquare}
    \sigma^2(m,z) = D_{+}^2(z)\int_{0}^{\infty} \frac{dk}{2\pi} \, k^2 P(k) W^2(k; R).
\end{eqnarray}
This is used as the input for the halo mass function from \cite{Tinker2008}:
\begin{eqnarray}
\label{eqn:hmfs:general_hmf}
    \frac{dn(m, z)}{dm} =  A\left[ \left(\frac{\sigma}{b}\right)^{-a} + 1\right] e^{ -c/\sigma^2} \frac{\bar{\rho}_{m,0}}{m} \frac{d \mathrm{ln}(\sigma^{-1})}{dm}.
\end{eqnarray}
where $\bar{\rho}_{m,0}$ is the mean density in the universe at
redshift $z=0$, $\delta_c \simeq 1.686$ is the critical overdensity
for collapse and $\lbrace A, a, b, c\rbrace$ are given the values
for $m_{200m}$ (the mass corresponding to the portion of the
cluster which has density greater than $200\bar{\rho}_{m,0}$) from
\cite{Tinker2008} of $\lbrace 0.186, 1.47, 2.57, 1.19\rbrace$.

When comparing to real-world clusters we need to correct for the
fact that theoretical mass functions are defined with respect to the
average matter density $\bar{\rho}_{m,z}$, but observers frequently
report cluster masses with respect to the critical density $\rho_c$.
In order to do this, we follow the procedures of
\cite{Waizmann2011a} and \cite{Mortonson2011} to convert all cluster
masses to  $m_{200m}$, and correct for Eddington bias.
Eddington bias refers to the fact that there is a larger population
of small mass haloes which may up-scatter into our observations than
there are high mass haloes which may down-scatter into them, and is
corrected for using:
\begin{eqnarray}
\label{eqn:edd}
\ln m^{Edd} = \ln m + \frac{1}{2}\epsilon \sigma_{\ln m}^{2},
\end{eqnarray}
where $\epsilon$ is the local slope of the halo mass function and
$\sigma_{\ln m}^{2}$ is the measurement uncertainty for the cluster
mass.

In order to ensure we are avoiding \textit{a posteriori} selection
(by only performing our test in regions in which we have already
observed something which we believe to be unusual) we set
$f_{sky}=1$. This is both the most conservative estimate and, we
believe, the correct one for testing `the most extreme clusters in
the sky'.
\subsection{Results}
\begin{figure}
 \begin{centering}
  \includegraphics[width=80mm]{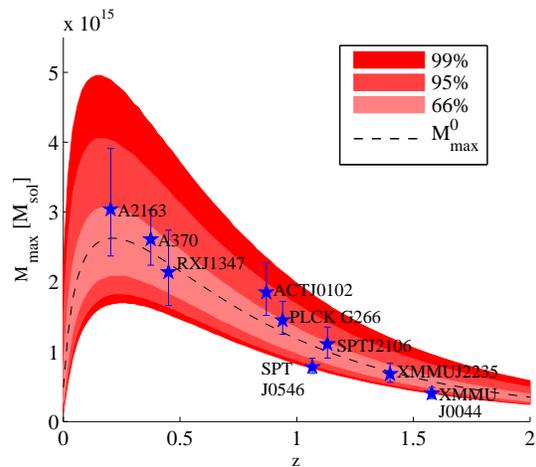}
 \caption{Extreme value contours and modal highest-mass cluster with redshift for a \lcdm cosmology, along with a set of currently observed `extreme' galaxy clusters. None lie in the region above the $99\%$ contour and hence are consistent with} a concordance cosmology.
    \label{fig:mmaxz}
 \end{centering}
\end{figure}
We now seek to use the apparatus described above to test if any
currently observed objects are significantly extreme to give us
cause to question \lcdm cosmology. We consider the set of recently
observed, potentially extreme clusters shown in table 1 in a \lcdm
cosmology as described above. The extreme value contours (light -
$99\%$, medium - $95\%$, dark - $66\%$), most likely maximum mass ${M}^{0}_{\rm max}$
(solid line) and the cluster masses and redshifts (stars) are
plotted in Figure \ref{fig:mmaxz}. The plot shows the expected
features of a peak in maximum halo mass at $z\approx0.2$ (the
location and height of which is in broad agreement with the analysis
of \citealt{Holz2010}). As can be seen, none of the currently
observed clusters lie outside the $99\%$ confidence regions of the
plot meaning that there is no current strong evidence for a need to
modify the \lcdm concordance model from high-mass high-redshift
clusters. This appears to be in agreement with the findings of
\cite{Waizmann2011a} for a similar set of clusters, but in
contradiction to \cite{Chongchitnan2011}\footnote{In an updated version of this analysis, Chongchitnan \& Silk find no tension with \lcdm} who find that the cluster
XMMUJ0044 is a $4\sigma$ result (i.e. should lie well above the
$99\%$ region in Figure \ref{fig:mmaxz}), whilst here we find it to
be well within the acceptable region.

\section{Testing Cosmological Models with Extreme Clusters}
\label{sec:codecs} In addition to simply ruling out \lcdm cosmology
with massive clusters, we may also consider whether extreme objects
offer the potential to discriminate between different alternative
models. Whilst many alternative models are capable of predicting
enhanced structure formation, the exact scale and time dependence of
the enhancement will differ from model to model. Here we consider
two models which have a well defined and investigated effect on the
halo mass function, and hence are relatively simple to calculate the
extreme value statistics over a range of redshift for: local form
primordial non-Gaussianity and the bouncing, coupled scalar field
dark energy model labelled as `SUGRA003' in \cite{Baldi2011}. These
should be regarded as toy models -- our aim is to show how the
extreme value statistics can be used to select between different
models, rather than make definite predictions.
\subsection{Models Considered}
We make use of the CoDECS simulations kindly made publicly available
by \cite{Baldi2010, Baldi2011a}. This suite of large N-body
simulations includes realisations of both \lcdm and a number of
coupled dark energy cosmologies. Here, we compare the CoDECS
$\Lambda$CDM-L (where `L' is for `Large') simulation of the
concordance cosmology to both the primordial non-Gaussianity and the
SUGRA003 (supergravity) bouncing dark energy models. Primordial
non-Gaussianity, motivated by considerations of the fluctuations of
the inflaton field, is one of the most widely explored modifications
to the concordance cosmology \citep[e.g.][]{Desjacques2010} and has
long \citep{Lucchin1988a} been known to affect the abundances of
high-mass galaxy clusters. It has also been the model most invoked
\citep{Jimenez2009, Cay'on2011, Chongchitnan2011, Hoyle2011b} to
account for apparently over-massive high redshift objects, all of
these authors reporting values of local non-Gaussianity parameter
$f_{NL} \sim 300-500$ as being able to account for such clusters.

However, \cite{Baldi2011} points out that such models enhance
numbers of high mass clusters at all redshifts, creating tension
with observations at low redshift in the attempt to alleviate them
at high redshift. As an alternative scenario, the
supergravity-motivated scalar field scenario of \cite{brax1999d} is
considered. This model includes a scalar field component $\phi$
which couples to dark matter with a coupling strength $\beta$ and
has the self interacting potential:
\begin{eqnarray}
V(\phi) = B\phi^{-\alpha}e^{\phi^2/2}
\end{eqnarray}
This scalar field component acts as a `bouncing' dark energy;
structure formation is enhanced at early times, but is suppressed
with respect to \lcdm after the point at which the evolution of
$\phi$ changes sign (the `bounce'), meaning \lcdm values for
$\sigma_8$ can still be reproduced at $z=0$. In order to match
background observables given by WMAP7 constraints
\citep{Komatsu2011n}, the SUGRA003 version of the potential has
`tuned' parameters $\lbrace B, \alpha, \beta \rbrace = \lbrace
0.0202, 2.15, -0.15 \rbrace$.

For \lcdm and SUGRA models, we fit a halo mass function of the
\cite{Tinker2008} form directly to the haloes identified using a
friends-of-friends (FoF) algorithm with linking length
$l=0.2\bar{d}$ (where $\bar{d}$ is the mean inter-particle
separation) in the relevant CoDECS simulation ($\Lambda$CDM-L and
SUGRA003-L respectively). For the primordial non-Gaussianity model,
we apply a non-Gaussian correction factor $\rnl$ to the halo mass
functions found in the $\Lambda$CDM-L simulation, choosing the
$\rnl$ of \cite{LoVerde2008} (LMSV):
\begin{eqnarray}
\label{eqn:hmf_evs:rlmsv}
\lefteqn{\mathcal{R}_{LMSV}(f_{NL}) =} \nonumber \\
& 1 + \frac{\sigma^2}{6\delta_c}\left[ S_3(\sigma) \left( \frac{\delta_c^4}{\sigma^4} - \frac{2\delta_c^2}{\sigma^2} - 1 \right) + \frac{d S_3}{d \ln\sigma}\left(\frac{\delta_c^2}{\sigma^2} - 1 \right) \right].
\end{eqnarray}
where $S_3$ is the normalised skewness of the matter density field,
for which we use the approximation:
\begin{eqnarray}
\label{eqn:hmf_evs:s3}
S_3 \simeq 3 \times 10^{-4} f_{NL} \sigma^{-1}
\end{eqnarray}
given by equation (2.7) of \cite{Enqvist2011}. We adopt a value of
$f_{NL}=300$ for our non-Gaussian model as it is both consistent
with the observational findings discussed above and leads to a
similar magnitude of enhancement of structure formation at high
redshifts as the SUGRA003 model.

The values of $H(z)$ and $D_{+}(z)$ required to find $dV/dz$ for all
three models are calculated using the tabulated growth functions and
expansion histories for the cosmologies, numerically calculated from
the evolution equations and provided on the CoDECS website.
\subsection{Results}
\begin{figure}
 \begin{centering}
  \includegraphics[width=80mm]{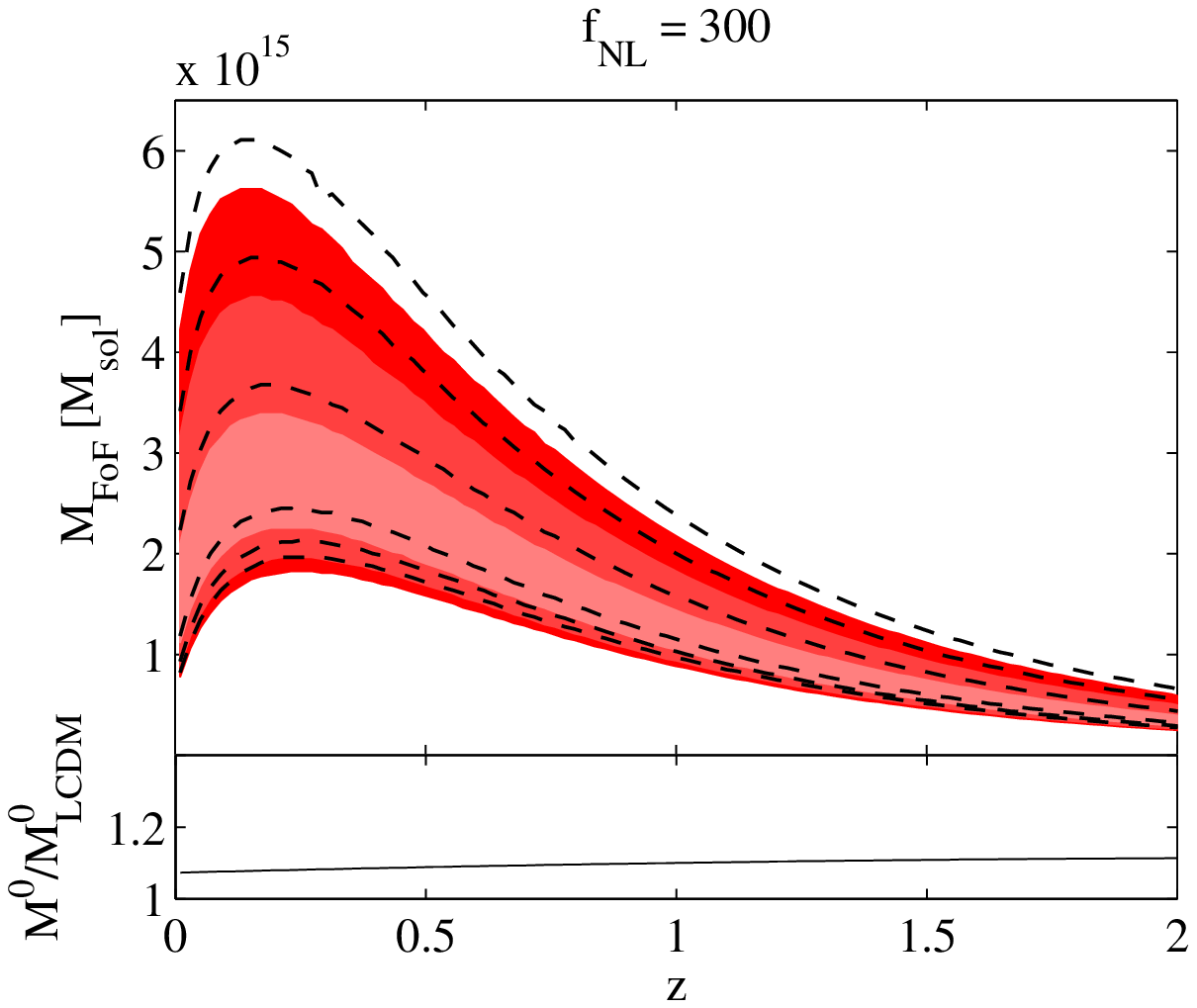}
  \includegraphics[width=80mm]{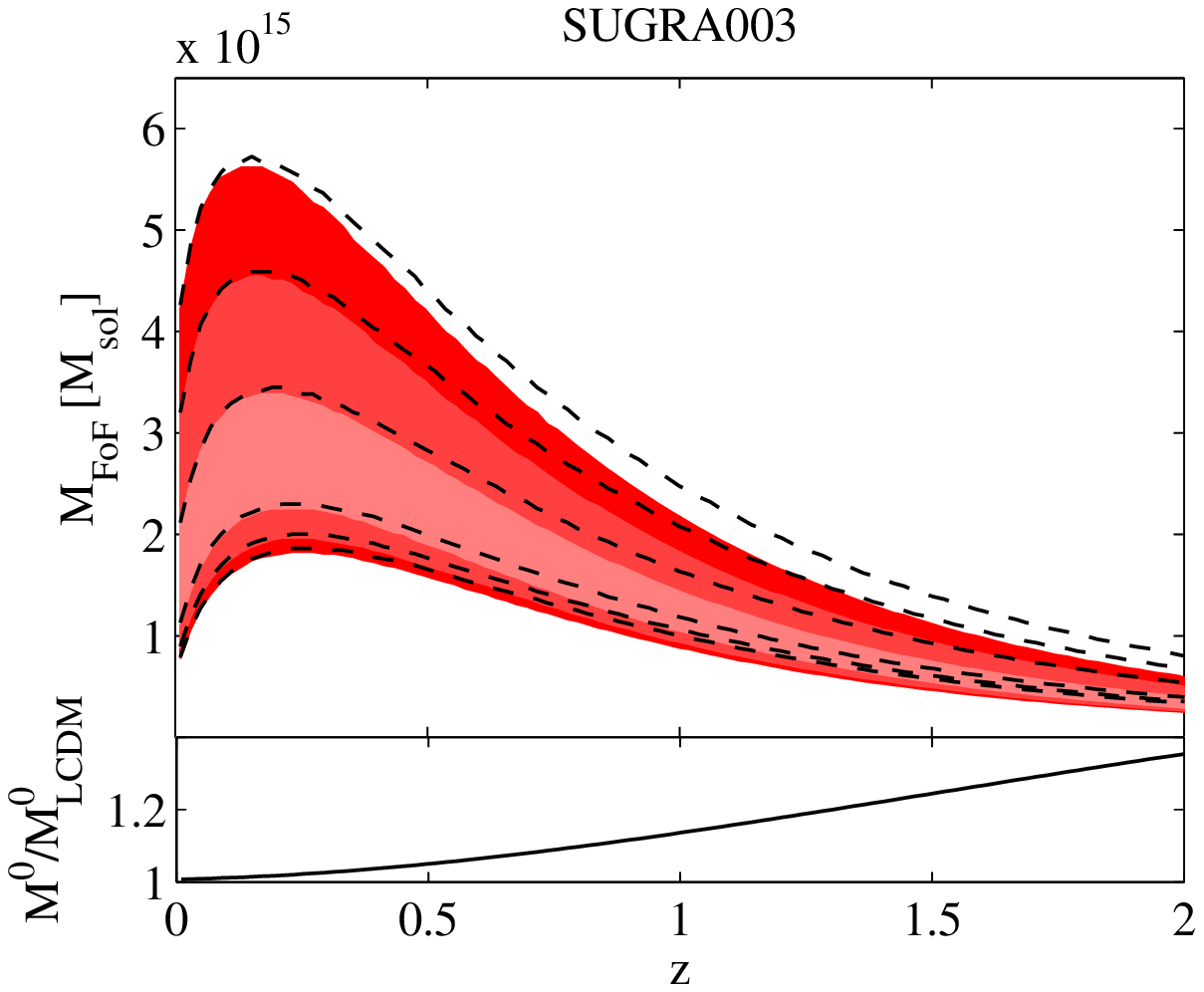}
 \caption{Extreme value contours for \lcdm (shaded regions), $f_{NL}$ and SUGRA models (dashed lines). Lower plots are the enhancement of modal highest-mass cluster over the \lcdm value, showing different behaviour for the two alternative models.}
    \label{fig:models}
 \end{centering}
\end{figure}
With the halo mass functions and expansion histories of each
cosmology we are then able to carry out the procedure of section
\ref{sec:methods} to find the EVS of objects within an
observational survey in each cosmology, the results of which are
shown in Figure \ref{fig:models}. Plotted are extreme value contours
(light - $99\%$, medium - $95\%$, dark - $66\%$) for the \lcdm model
and the edges of the three extreme value contours for the
non-Gaussian and SUGRA models (dashed lines) as well as the
enhancement in the most likely maximum mass $M^0_{\rm max}$ over the \lcdm
predictions. As can be seen (and as expected) the primordial
non-Gaussianity model shows an enhancement of the mass of the
highest mass cluster at all redshifts, whilst the SUGRA model is
capable of enhancing $M^0_{\rm max}$ at high redshifts whilst leaving
it unchanged at more recent times. Thus, if \lcdm is ruled out by
both high and low redshift clusters, primordial non-Gaussianity
could be seen as the favoured explanation whilst, if only high
redshift observations appear in contradiction, both non-Gaussian and
SUGRA models would be allowed (unless the limit of an ideal, complete survey was reached).
\section{Discussion and Conclusions}
\label{sec:conclusions}
In this paper we have presented a theoretical framework for the interpretation of extremely massive clusters in cosmological surveys. We avoid ambiguities in the volume being probed by observed clusters which occur in the alternative approach of using single clusters to estimate the halo mass function $n(m)$. Our method, in contrast, does not require such normalisation and is both more \textit{robust} and, inevitably, more \textit{conservative}, a combination of attributes which we believe should best describe confrontations of theory with observations.

We have used this method to show that existing observations of extreme clusters are not incompatible with the predictions of \lcdm concordance cosmology. However in anticipation of future, more complete surveys, we have also presented a definitive calculation of the region in the mass-redshift plane in which the presence of a single cluster would exclude the standard model with high confidence. We have also shown that different cosmological models including physics beyond the concordance picture are capable of making significantly different predictions of the allowable regions in the mass-redshift plane. This means that whilst current extreme cluster data do not discriminate between the \lcdm model and these alternatives, future data may well be capable of doing so. We intend in due course to extend the approach presented here by incorporating it into a full model selection analysis.
\section*{Acknowledgments}
Ian Harrison receives funding from an STFC studentship and would
like to thank Geraint Pratten and Marco Baldi for useful
discussions. For the purposes of this work Peter Coles is supported
by STFC Rolling Grant ST/H001530/1.

\end{document}